\documentclass[a4paper]{raa}            
\usepackage{CJK}
\usepackage{graphicx,times}             
\usepackage{natbib}
\usepackage{amssymb,amsmath}
\bibpunct{(}{)}{;}{a}{}{,}

\usepackage[a4paper=true,pagebackref=true]{hyperref}
\hypersetup{colorlinks = true, linkcolor = blue, anchorcolor = red, citecolor = blue, filecolor = red, pagecolor = red, urlcolor = red}

\usepackage{multirow}
\begin{document}
\begin{CJK*}{UTF8}{gbsn}

    \title{The Metallicity Dimension of the Super Earth--Cold Jupiter Correlation}
    
   \volnopage{Vol.0 (20xx) No.0, 000--000}      
   \setcounter{page}{1}          

   \author{Wei Zhu (祝伟)\inst{1}}

   \institute{Department of Astronomy, Tsinghua University, Beijing 100084, China; {\it weizhu@tsinghua.edu.cn} \\
\vs\no
   {\small Received~~20xx month day; accepted~~20xx~~month day}}

\abstract
{The correlation between close-in super Earths and distant cold Jupiters in planetary systems has important implications for their formation and evolution. In contrary to some earlier findings, a recent study conducted by Bonomo et al.\ suggests that the occurrence of cold Jupiter companions is not excessive in super Earth systems. Here we show that this discrepancy can be seen as a Simpson's paradox and is resolved once the metallicity dependence of the super Earth--cold Jupiter relation is taken into account. A common feature is noticed that almost all the cold Jupiter detections with inner super Earth companions are found around metal-rich stars. Focusing on the Sun-like hosts with super-solar metallicities, we show that the frequency of cold Jupiters conditioned on the presence of inner super Earths is $39_{-11}^{+12}\%$, whereas the frequency of cold Jupiters in the same metallicity range is no more than $20\%$. Therefore, the occurrences of close-in super Earths and distant cold Jupiters appear correlated around metal-rich hosts. The relation between the two types of planets remains unclear for stars with metal-poor hosts due to the limited sample size and the much lower occurrence rate of cold Jupiters, but a correlation between the two cannot be ruled out.
\keywords{Planetary systems -- Stars: abundances -- Methods: statistical}
}

   \authorrunning{Wei Zhu }            
   \titlerunning{The metallicity dimension of the super Earth--cold Jupiter correlation}  

   \maketitle
%

\section{Introduction}

Planets with masses/radii between Earth and Neptune are abundant in the Galaxy, with each system expected to have several of these so-called super Earths within the inner $\sim1\,$au \citep[e.g.,][]{Mayor:2011, Petigura:2013, Fressin:2013, Zhu:2018,Mulders:2018, Hsu:2019}. Consequently, these systems contain significantly more solid material compared to the Solar system. If the planets in these systems were formed \textit{in situ}, then the planet-forming discs might have been (or near) gravitationally unstable \citep{Chiang:2013,Schlichting:2014}. As a result, theories involving disk-driven migration of (proto-) planets and/or planet-forming material have been popular.

To further constrain the formation models, the presence/absence of cold Jupiter companions (with semi-major axis $\gtrsim1\,$au and mass or minimum mass $>0.3\,M_{\rm J}$) in systems with inner super Earths has been considered a useful test \citep[e.g.,][]{IdaLin:2010,Izidoro:2015}. Based on carefully constructed samples of radial velocity (RV) and transiting systems, \citet{ZhuWu:2018} showed that cold Jupiters were found in roughly one third of super Earth systems, outnumbering the frequency of cold Jupiters around field stars with similar properties by a factor of $\sim3$. This excess of cold Jupiters was subsequently confirmed by several other studies \citep{Bryan:2019,Rosenthal:2022}. It is also in alignment with the excess of long-period transiting events in Kepler transiting systems, if these transiting events are indeed due to genuinely cold Jupiters \citep{Foreman-Mackey:2016, Herman:2019, Masuda:2020}. The inversion of this conditional rate seems to suggest that most, if not all, of cold Jupiter systems should have inner super Earths (\citealt{ZhuWu:2018, Bryan:2019}; but see also \citealt{Barbato:2018} and \citealt{Rosenthal:2022} for some lower estimates). The observed strong correlation between super Earths and cold Jupiters has motivated further development of various theoretical models \citep[e.g.,][]{Chen:2020, Bitsch:2020, Schlecker:2021,Bitsch:2023,Chachan:2023,Best:2023}.

A recent study by \citet{Bonomo:2023} arrived at an opposite conclusion. Based on radial velocity observations from the HARPS-N survey of 37 Kepler and K2 systems with close-in ($P<100$\,d) super Earths,
\footnote{We follow \citet{Bonomo:2023} and also exclude Kepler-22 from the analysis. With a host metallicity of [Fe/H]$=-0.26$, this has no impact on the conditional rate that is derived in Section~\ref{sec:correlation}.}
\citet{Bonomo:2023} reported the detections of five cold Jupiters and two potential cold Jupiter candidates in four systems. Statistical analysis led to an occurrence rate of $9.3^{+7.7}_{-2.9}\%$ for cold Jupiters between 1--10 au around systems with inner super Earths. According to \citet{Bonomo:2023}, this fraction is lower than the fraction of similar cold Jupiters around field stars reported in \citet{Wittenmyer:2020}. Although their sample remains too small to draw any firm conclusion, it hints at an anti-correlation between small planets and cold Jupiters.

\begin{figure*}
    \centering
    \includegraphics[width=\textwidth]{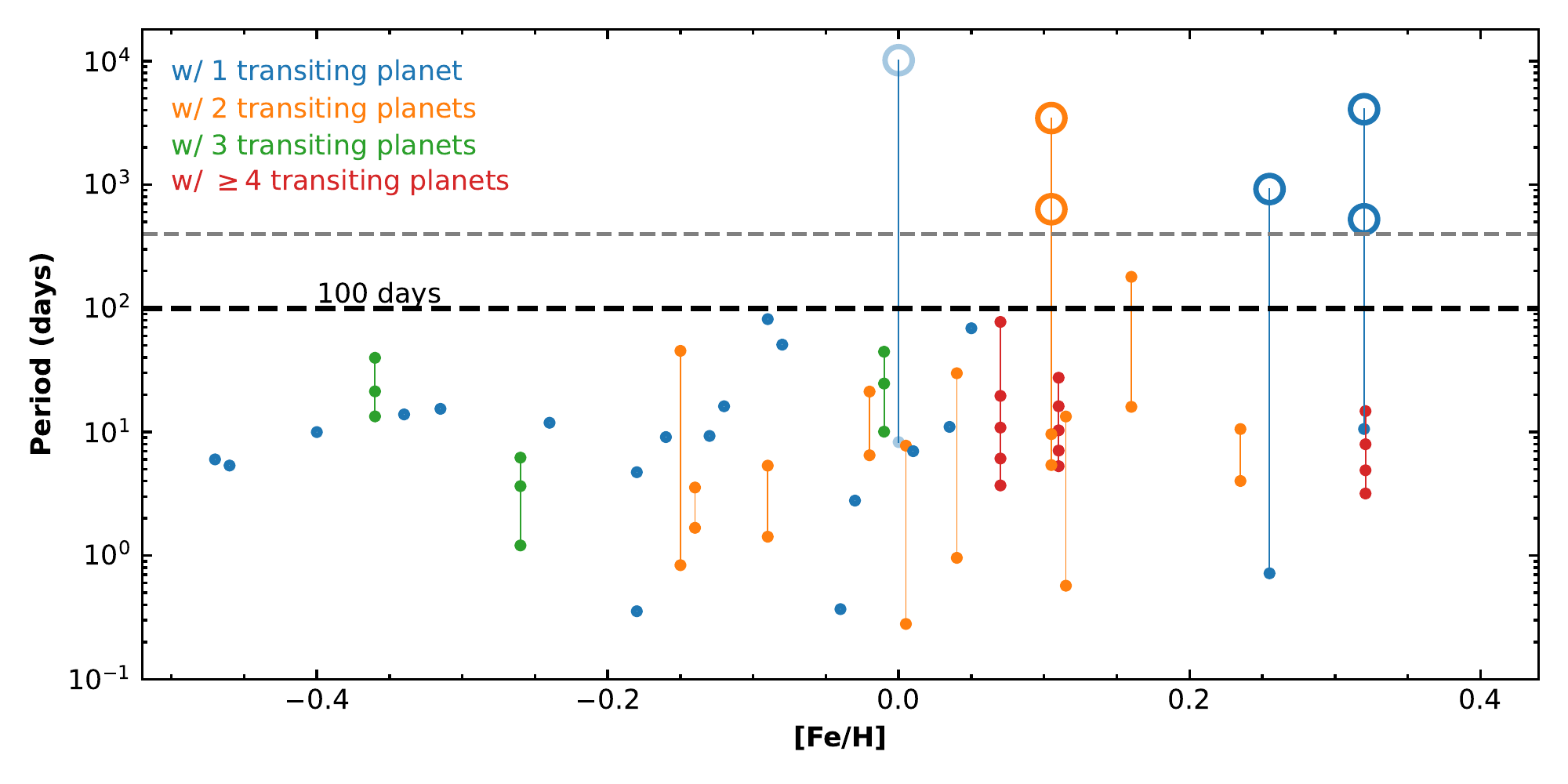}
    \caption{Illustration of the planetary systems observed by HARPS-N. The $x$-axis shows the bulk metallicity of the host star, and the $y$-axis shows the orbital period of the detected planets. Systems with different observed transit multiplicities have been differentiated with different colors, and the detected cold Jupiters (including the tentative detection in K2-12 at [Fe/H]$=0$) are shown with open circles. For the purpose of illustrations, systems with nearly the same host metallicities have been offset by $\pm0.005$ in [Fe/H].}
    \label{fig:sample}
\end{figure*}

Putting the discrepancy on the overall conditional rate aside, it is interesting to note that all the cold Jupiter detections are from host stars that are relatively metal-rich. The average metallicity of the HARPS-N sample is ${\rm [Fe/H]}\approx -0.05$, but for simplicity we will set the threshold at ${\rm [Fe/H]}=0$, i.e., (nearly) solar metallicity. As illustrated in Figure~\ref{fig:sample}, the bulk metallicities of the three cold Jupiter hosts are ${\rm [Fe/H]}=0.11\pm0.06$, $0.255\pm0.065$, and $0.32\pm0.08$ for Kepler-68, K2-312/HD 80653, and Kepler-454, respectively, and another system (K2-12) with a candidate companion that remains consistent with a cold Jupiter within 10\,au has [Fe/H]$=0.00\pm0.07$. This is a striking feature given that only $35\%$ (13/37) of the selected systems have [Fe/H]$>0$. In other words, the derived conditional rate could be nearly three times higher if only super-solar metallicity stars are considered.

The feature that cold Jupiters with inner small planet companions are preferentially (or exclusively) found around metal-rich stars is already seen in the original samples used by \citet{ZhuWu:2018} and \citet{Bryan:2019} and also seen in the recent samples of \citet{Rosenthal:2021}, \citet{VanZandt:2023}, and \citet{Weiss:2023}. In particular, the pure-RV sample used in \citet{ZhuWu:2018} and tabulated in their table~2 shows that cold Jupiters were not detected in super Earth systems with ${\rm [Fe/H]}\lesssim0.1$. As for the transiting systems, all those with cold Jupiter detections in resolved orbits have stellar ${\rm [Fe/H]}>0$ \citep{ZhuWu:2018, Bryan:2019,VanZandt:2023,Weiss:2023}. See the Figure~8 of \citet{ZhuDong:2021} for a similar illustration to Figure~\ref{fig:sample}. There were two metal-poor systems (Kepler-93 and Kepler-97) which showed long-term RV trends and could have been explained by distant cold Jupiters, but the more recent RV observations have ruled out the cold Jupiter explanation \citep{Weiss:2023,Bonomo:2023}.

Observational bias is unlikely the reason for the observed feature, because cold Jupiters around metal-poor stars can be and have been detected. According to a query made on June 26, 2023 to the NASA Exoplanet Archive \citep{Akeson:2013}, there have been $\sim130$ cold Jupiters around Sun-like stars, and $>30$ have host metallicity ${\rm [Fe/H]}<0$. Therefore, it seems to be the case that the occurrence of cold Jupiters in systems with inner small planets must correlate with the host star metallicity, a point that has been made in \citet{ZhuWu:2018}. According to that study, the rate of cold Jupiters conditional on the presence of inner super Earths, $P({\rm CJ}|{\rm SE})$, increases to $\sim50\%$ for systems with bulk metallicities above the solar value, whereas it averages to $\sim30\%$ if the whole dynamical range of stellar metallicities are considered. This is not much a surprise, given that the occurrence of giant planets in general correlates with the host star metalicity \citep[e.g.,][]{Santos:2001, Santos:2004,Fischer:2005} and that if indeed the majority of the cold Jupiters have inner small companions \citep{ZhuWu:2018, Bryan:2019}. However, whether or not the metallicity dependence of the super Earth--cold Jupiter correlation is entirely due to the giant planet--metallicity correlation requires further investigations with much larger planet samples.

In this work, we show that the discrepancy between \citet{Bonomo:2023} and previous works can be resolved by adding the metallicity dimension of the super Earth--cold Jupiter correlation.
Specifically, the super Earth systems with metal-rich (${\rm [Fe/H]}>0$) hosts have an excess of cold Jupiters, whereas it is unclear for systems with metal-poor hosts, but a correlation between the two planet populations cannot be ruled out.

\section{Excess of Cold Jupiters in metal-rich systems} \label{sec:correlation}

\begin{table*}
\caption{The conditional rate $P({\rm CJ}|{\rm SE, [Fe/H]>0})$ derived from different samples.}
\label{tab:rates}
\centering
\begin{tabular}{lccc}
\hline\hline
Sample & $P({\rm CJ}|{\rm SE,\,[Fe/H]>0})$ & Label & Note \\
Kepler \& K2 sample ($R_{\rm SE}<4\,R_\oplus$) & 3/13 ($29_{-11}^{+13}\%$) & B23 & (1) \\
Kepler \& K2 sample ($R_{\rm SE}<4\,R_\oplus$, w/o transit compact multi) & 3/10 ($36_{-14}^{+15}\%$) & B23, clean & (2)\\
RV sample ($m_{\rm SE}<20\,M_\oplus$) & 9/17 ($53\pm11\%$) & RV, high & (3) \\
RV sample ($m_{\rm SE}<10\,M_\oplus$) & 3/7 ($44_{-16}^{+17}\%$) & RV, low & (4) \\
RV+Transit & $39_{-11}^{+12}\%$ & Joint & (5) \\
\hline
\end{tabular} \\
The ratios give the numbers of systems with both CJ and SE out of the numbers of systems with SEs. The fractions in the parentheses are the estimated rates with uncertainties. \\
\footnotemark[1]{From the sample of \citet{Bonomo:2023}. The rate is calculated by adopting a mean efficiency of $87.9\%$.} \\
\footnotemark[2]{Similar to (1), but now with the transit compact multis (i.e., systems with at least four transiting planets) removed.} \\
\footnotemark[3]{From the RV sample of \citet{ZhuWu:2018}. Systems with at least one inner planet whose mass is below $20\,M_\oplus$ are considered. The rate is calculated by adopting 100\% detection efficiency.} \\
\footnotemark[4]{From the RV sample of \citet{Bryan:2019} after the removal of systems with M-dwarf hosts. Systems with at least one inner planet whose mass is below $10\,M_\oplus$ are considered. The rate is calculated by adopting 100\% detection efficiency.} \\
\footnotemark[5]{The joint constraint on the conditional rate, derived from combination of the ``B23, clean'' and ``RV, low'' samples.}
\end{table*}

\begin{figure}
\centering
\includegraphics[width=0.7\columnwidth]{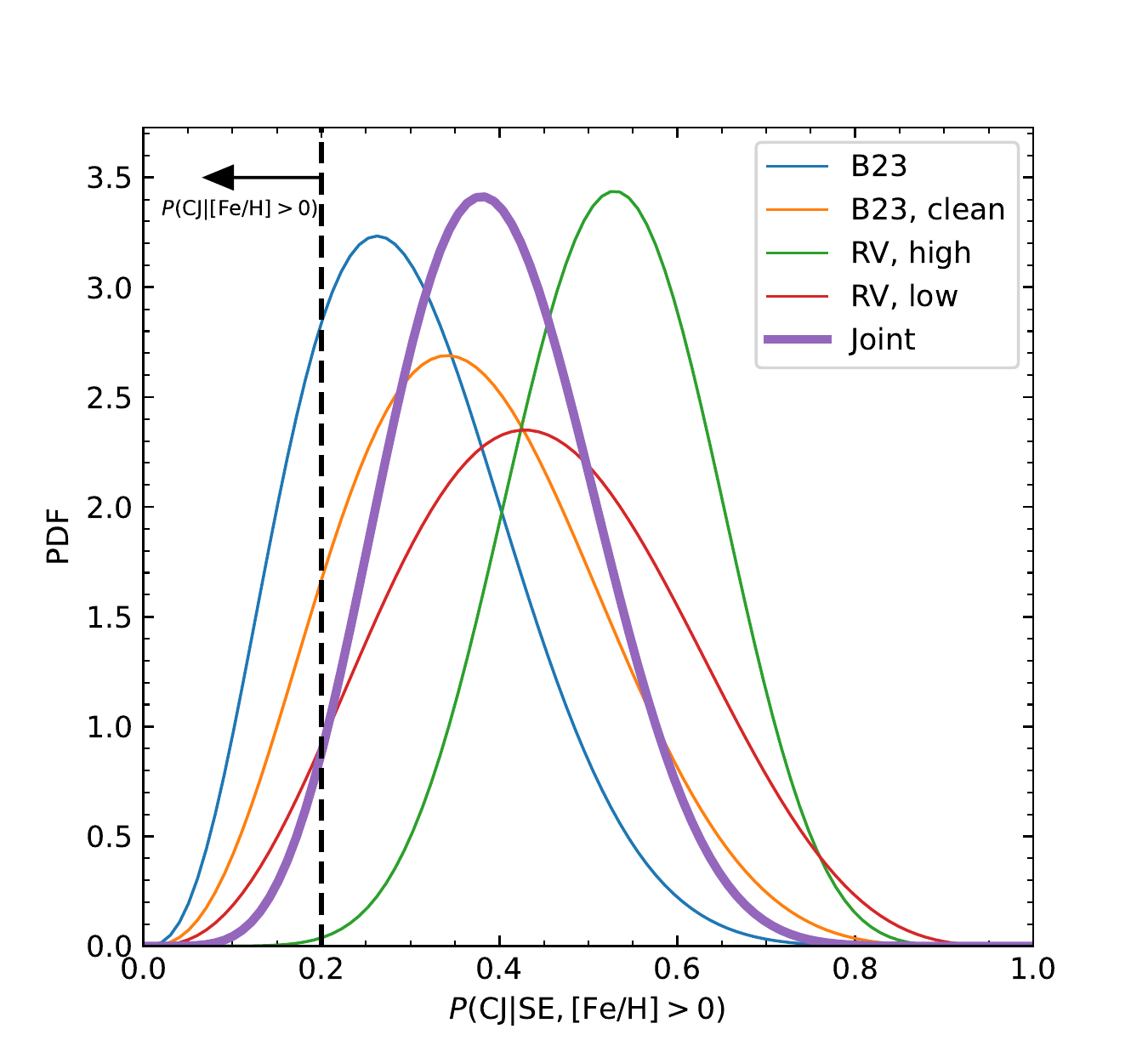}
\caption{Illustration of the conditional rate, $P({\rm CJ}|{\rm SE, [Fe/H]>0})$, derived from different samples, in comparison to the estimated upper limit of the rate $P({\rm CJ|[Fe/H]>0})$. See Table~\ref{tab:rates} for the meanings of the labels.}
\label{fig:rates}
\end{figure}

\subsection{The conditional rate $P({\rm CJ}|{\rm SE, [Fe/H]>0})$}

We first estimate the conditional rate $P({\rm CJ}|{\rm SE})$ in the super-solar metallicity regime, which is denoted $P({\rm CJ}|{\rm SE, [Fe/H]>0})$. There are 13 planetary systems with [Fe/H] $>0$ in the 37 HARPS-N systems of \citet{Bonomo:2023}.
\footnote{Two systems have reported [Fe/H]$=0$, including one (K2-12) which shows a linear trend consistent with a cold Jupiter within 10\,au. Including or not these two systems makes no statistically significant difference in the final result. We therefore decide to exclude them to avoid discussing the ambiguity on the nature of the linear trend in K2-12.}
As shown in Figure~\ref{fig:sample}, this subset of the sample contains all of the confirmed cold Jupiter detections. According to \citet{Bonomo:2023}, the HARPS-N survey has an average completeness of $87.9\%$ for cold Jupiters within 10\,au. With these numbers, we can estimate the conditional rate $P({\rm CJ}|{\rm SE, [Fe/H]>0})$. We assume binomial statistics and adopt a flat prior on the conditional rate. With $n_s$ detections and $n_f$ non-detections, the postior distribution of the conditional rate is given by a beta distribution with shape parameters $a=n_s+1$ and $b=n_f+1$
\begin{equation}
    f(p; a, b) = \frac{1}{{\rm B}(a, b)} p^{a-1} (1-p)^{b-1}. 
\end{equation}
Here ${\rm B}(a, b)$ is the beta function, and we report the median value and the $68\%$ confidence interval centered on the median as the uncertainties, which can be efficiently computed via the \texttt{scipy.stats.beta} function \citep{scipy}. In the present case, we have $n_s=3$, $n_f=13\times 87.9\%-3$, and a conditional rate $P({\rm CJ}|{\rm SE, [Fe/H]>0})=29_{-11}^{+13}\%$. This is a factor of $\sim3$ higher compared to the conditional rate of \citet{Bonomo:2023} over the full metallicity range. We list this new conditional rate in Table~\ref{tab:rates}.

The HARPS-N sample has an over-representation of transiting systems with dynamically compact configurations (i.e., transiting compact multis), which also has implications to a proper derivation of the conditional rate. If we define a transiting compact multi as a system with at least four transiting planets within 100\,d, then the HARPS-N sample contains three such systems, all with ${\rm [Fe/H]}>0$. See Figure~\ref{fig:sample} for an illustration. The fraction of transiting compact multis in the super-solar regime, 3/9 (excluding K2 systems, which have much shorter time baseline, \citealt{Howell:2014}), is much higher than the fraction of transiting compact multis in the overall Kepler systems ($<3\%$, \citealt{Zhu:2018,Zink:2019, He:2020,ZhuDong:2021}). This latter fraction would be further reduced if one only selects Kepler systems with super-solar metallicities, because the transiting compact multis are preferentially found around metal-poor stars \citep[e.g.,][]{Brewer:2018,Anderson:2021}.
The selection of the HARPS-N sample is therefore biased towards transiting compact multis.

Theoretical studies have shown that the transiting compact multis can be easily dynamically perturbed by one or more distant massive companions, unless these companions are in very special configurations \citep[e.g.,][]{Carrera:2016,Becker:2017,Hansen:2017,Huang:2017,Lai:2017,Pu:2021}. It remains unclear whether these transiting compact multis follow the same correlation with cold Jupiters as the general super Earth population, and thus the bias towards the selection of transiting compact multis may imply a bias against the detection of cold Jupiters in such systems.
\footnote{In fact, the absence of Kepler compact multis around metal-rich stars has been used by \citet{ZhuWu:2018} as a supporting evidence for the super Earth--cold Jupiter correlation.}
Given this and the fact that these transiting compact multis are over-represented, it is reasonable to remove the three dynamically compact systems in the derivation of the conditional rate. By doing so, we are arrived at a slightly higher, but yet statistically consistent, rate of $36_{-14}^{+15}\%$, as also listed in Table~\ref{tab:rates}.

For comparisons, we also derive the same conditional rate from the pure-RV sample, in which all planets are detected via the RV method. This sample has not been expanded so much as the transit sample since the studies of \citet{ZhuWu:2018} and \citet{Bryan:2019}. For example, there is only one new system (HD 168009) with Sun-like hosts and low-mass ($<10\,M_\oplus$) close-in planets in the recently released California Legacy Survey \citep{Rosenthal:2021}. We therefore rely on the pure-RV samples of \citet{ZhuWu:2018} and \citet{Bryan:2019}. If we define a super Earth by its mass $m_{\rm SE}<20\,M_\oplus$, then the sample of \citet{ZhuWu:2018} is more suitable. According to the Table~2 of that study, there are nine systems with cold Jupiter companions out of a total of 17 super Earth systems with [Fe/H]$>0$, resulting in a conditional rate of $53\pm11\%$. If we adopt a lower mass limit (i.e., $10\,M_\oplus$) for super Earth and use the sample of \citet{Bryan:2019}, the numbers are three and seven, and the resulting conditional rate is $44_{-16}^{+17}\%$. In the above discussion we have limited ourselves to planetary systems with Sun-like hosts, defined by stellar mass in the range 0.7--1.3\,$M_\odot$, as the planet distribution seems to vary with host star types (see \citealt{ZhuDong:2021} and references therein). This excludes 23 of the 65 systems in \citet{Bryan:2019}, making their sample comparable in size with that of \citet{ZhuWu:2018}. We have also adopted a 100\% detection efficiency for the detection of cold Jupiters in such pure-RV systems. This is shown to be reasonable according to the efficiency estimation of \citet[][see their Figure~5]{Bryan:2019}.

As listed in Table~\ref{tab:rates} and illustrated in Figure~\ref{fig:rates}, the conditional rates of cold Jupiters around Sun-like stars with super-solar metallicities, derived from both transit and pure-RV samples under different criteria, appear broadly consistent with each other. When the two independent samples, namely the HARPS-N sample without the transiting compact multis and the pure-RV sample with the lower mass limit of super Earths, are combined, we find a conditional rate of $39_{-11}^{+12}\%$. In other words, $\sim40\%$ of Sun-like stars with inner super Earths and super-solar metallicities should have cold Jupiters in the distance range of 1--10\,au.

\subsection{On the frequency of cold Jupiters}

Is the conditional rate derived above higher than the rate of cold Jupiters in the same metallicity range, namely $P({\rm CJ|[Fe/H]>0})$? A detailed derivation of this rate is beyond the scope of the current paper. Instead, we provide a reasonable estimate based on a simple approach. According to \citet{Cumming:2008} and as further explained in \citet{ZhuWu:2018}, the cold Jupiter rate, integrated over the full metallicity range, is $\sim 10\%$. The average metallicity of stars in the RV survey is usually around or slightly below the solar value \citep[e.g.,][]{Cumming:2008, Johnson:2010, Mayor:2011}. Therefore, even in the most extreme case that the more metal-rich half of the stars contribute all the cold Jupiter detections, the frequency of such planets around Sun-like stars with super-solar metallicities is $20\%$ at most. More realistic giant planet--metallicity correlations would give lower values. For example, the giant planet--metallicity correlation of \citet{Johnson:2010} indicates that the frequency of giant planets with [Fe/H]$>0$ is only $\sim1.4$ times the frequency of giant planets across the whole metallicity range. Therefore, the rate of cold Jupiters around metal-rich Sun-like stars is statistically lower than the conditional rate of cold Jupiters in the same metallicity range.

The recent California Legacy survey (CLS) provided another estimate of the cold Jupiter rate \citep{Rosenthal:2021,Fulton:2021}. By focusing on the Sun-like sample and applying a proper method that takes into account the impact of planet multiplicity, \citet{Zhu:2022} reported a rate of $17\%$ for the frequency of planetary systems containing at least one cold Jupiter around all metallicities. However, one should be cautious in directly applying this value to the current context. Some fraction of the cold Jupiters have inner giant companions, as inferred from the RV follow-up observations of close-in giant planets (namely hot and warm Jupiters, \citealt{Knutson:2014, Bryan:2016}), whereas none of the planetary systems in the \citet{Bonomo:2023} sample contains such close-in giants. This difference usually has negligible impact on the correlation study, as the frequency of close-in giants is typically small compared to the frequency of cold Jupiters. This is not the case for the CLS Sun-like sample, whose frequency of close-in giants is surprisingly high at $7.2\%$ \citep{Zhu:2022}. In particular, the hot Jupiter rate of 2.8\% is almost three times the canonical value ($\sim$1\%) from previous studies \citep[e.g.,][]{Cumming:2008, Mayor:2011}. In order to have a fair comparison, one should exclude the cold Jupiters with close-in giant companions from the overall cold Jupiter population. By doing so, we get $P({\rm CJ})$ between $\sim10\%$ (if all close-in giants have cold Jupiter companions) and $\sim13\%$ (if half of the close-in giants have cold Jupiter companions). These values are in a closer agreement with the adopted value of $10\%$.

Out of all 474 stars in the CLS Sun-like sample, $59\%$ have [Fe/H]$>0$ and contribute 41 out of 49 systems with at least one cold Jupiter detection. These values imply that the cold Jupiter frequency around stars with super-solar metallicities is 1.4 times the overall cold Jupiter frequency. This factor is consistent with our estimation based on the giant planet--metallicity correlation of \citet{Johnson:2010}. Therefore, the value of $P({\rm CJ|[Fe/H]>0})$ is probably in the range 14--19\%, which is below the upper limit of 20\% that we have adopted.

Finally, we comment on the unconditional rate, $P({\rm CJ})$, used in \citet{Bonomo:2023}. \citet{Bonomo:2023} quoted 20.2\% by summing up the occurrence rates of cold Jupiters from 300 to 10,000\,d, using the occurrence rates of gas giants from \citet{Wittenmyer:2020}. The above orbital period range does not match exactly the semi-major axis range of 1--10\,au, and the integrated occurrence rate does not account for the impact of giant planet multiplicity. A closer examination of the planet sample from \citet{Wittenmyer:2020} reveals that half of the cold Jupiter detections in the period range of 300--1000\,d have $P<400\,$d, including five with period below one year (although only two with derived semi-major axis below $1\,$au). If we adopt the lower limit in orbital period at 1\,yr, then the integrated occurrence rate in the period range 1\,yr$<P<10,000$\,d ($\sim9\,$au) becomes $16.2\%$. The revision of the upper limit to $10\,$au does not further revise this value. Furthermore, in order to obtain the frequency of cold Jupiter systems, it is necessary to correct for planet multiplicity, as a significant fraction of cold Jupiters reside in systems with multiple cold Jupiters. Based on the estimated average multiplicity of $1.27$ for cold Jupiters in the range of 1--10\,au \citep{Zhu:2022},
\footnote{As has been discussed above, the estimated rates of cold Jupiters in \citet{Zhu:2022} might suffer from systematics in the sample selection. However, the average multiplicity, which is the ratio between the frequency of planets and the frequency of planetary systems, should be less affected.}
we obtain an unconditional rate of $12.8\%$. Given the statistical uncertainty and probably systematic uncertainty arising from the use of the inverse detection efficiency method (see Section~1.2 of \citealt{ZhuDong:2021} for further discussions), it is not significantly different from the value used in this study for the full metallicity range (i.e., 10\%).

\section{Discussion} \label{sec:conclusion}

The coexistence (or not) of inner super Earths and outer cold Jupiters in planetary systems has significant theoretical implications for their formation and evolution. 
Several studies have shown that Sun-like stars with super Earths are more likely to have cold Jupiter companions compared to random field stars, suggesting that these two types of planets indeed tend to coexist. 

A recent study by \citet{Bonomo:2023} conducted RV follow-up observations of a sample of 37 planetary systems with inner super Earths (with periods less than 100\,d) and found that this sample of systems did not show an excess of cold Jupiter detections. It is first noticed that all the cold Jupiter detections in their sample are found around metal-rich stars (see Figure~\ref{fig:sample}). This feature is also seen in almost all the other known cold Jupiter systems with inner super Earth companions \citep{ZhuWu:2018,Bryan:2019,Rosenthal:2021,VanZandt:2023,Weiss:2023}. The study of the correlation (or not) between inner small planets and outer cold giants should take the stellar metallicity dependence into account.

This work derives $P({\rm CJ}|{\rm SE, [Fe/H]>0})$, the frequency of cold Jupiters around Sun-like stars conditioned on the presence of inner super Earths and super-solar metallicities. For the HARPS-N sample of \citet{Bonomo:2023}, this quantity is estimated to be $36_{-14}^{+15}\%$ after the removal of the transiting systems with dynamically compact configurations, on the basis that such systems are over-represented in the sample and potentially bias against the detection of cold Jupiters. The derived conditional rate is statistically consistent with the rate derived from the pure-RV sample. Combining the two samples, we estimate that $39_{-11}^{+12}\%$ of Sun-like stars with inner super Earths and bulk metallicity ${\rm [Fe/H]}>0$ should have cold Jupiter companions in the range of 1--10\,au. For comparisons, the frequency of cold Jupiters in the same metallicity range is $\lesssim 20\%$. Therefore, the current results support that super Earth systems around metal-rich hosts have an excess of cold Jupiters.

Almost no cold Jupiter has been detected in super Earth systems with Sun-like stars and sub-solar metallicities (i.e., ${\rm [Fe/H]}<0$).
\footnote{According to NASA Exoplanet Archive \citep{Akeson:2013}, HD 137496 and HD 191939 are perhaps the only exceptions. The former contains one confirmed cold Jupiter and have slightly sub-solar metallicity (${\rm [Fe/H]}=-0.027\pm0.040$, \citealt{AzevedoSilva:2022}). The latter has ${\rm [Fe/H]}=-0.15\pm0.06$ and a long-term RV trend that is consistent with a super-Jupiter beyond $3\,$au \citep{OrellMiquel:2023}, but more observations are needed to confirm the signal.}
Does this mean that there is no correlation or even an anti-correlation between inner super Earths and outer cold Jupiters in the metal-poor environment? Unfortunately we cannot answer this question with the currently available data. The frequency of cold Jupiters around metal-poor stars is intrinsically low, likely at the few percent level, so a much larger planet sample is needed to address the super Earth--cold Jupiter relation in this metallicity range. For example, using the HARPS-N sample of \citet{Bonomo:2023}, which contains 22 stars and no cold Jupiter detections with ${\rm [Fe/H]}<0$, one find the $95\%$ upper limit on the conditional rate to be $P({\rm CJ}|{\rm SE, [Fe/H]<0})<14\%$ following the same procedure as in Section~\ref{sec:correlation}. Either correlation or anti-correlation remains possible.

In both the metal-rich and metal-poor regimes, the frequency of cold Jupiters conditioned on the presence of super Earths appear consistent between the HARPS-N sample of \citet{Bonomo:2023} and the original samples of \citet{ZhuWu:2018} and \citet{Bryan:2019}. Therefore, the discrepancy on the overall conditional rate can be explained by Simpson's paradox: the same trend appears in different subsets of the data, but disappears or even reverses in the joint dataset.

The current planet sample remains too small to fully establish the correlation between inner small planets and outer giant planets and explore its dependence on properties of the host stars and the planetary systems (see some early attempts by \citealt{Zhu:2019} and \citealt{He:2023}), making it difficult to test predictions of formation models \citep[e.g.,][]{Schlecker:2021,Bitsch:2023, Best:2023}. The ongoing RV follow-up observations of transiting systems \citep[e.g.,][]{VanZandt:2023,Weiss:2023,Bonomo:2023} and upcoming Gaia astrometric detections \citep[e.g.,][]{Perryman:2014, EspinozaRetamal:2023} aided by careful statistical analysis will be helpful in this regard. Further studies of this correlation can also be a task for future missions \citep[e.g.,][]{Ji:2022, Ge:2022, Wang:2023}.

\begin{acknowledgements}
We would like to thank the anonymous referees for critics and comments that have improved the quality of this work.
We thank Bert Bitsch, Aldo Bonomo, Subo Dong, and Yanqin Wu for discussions and comments on some earlier version of the manuscript.
This work is supported by the National Science Foundation of China (grant No.\ 12173021 and No. 12133005) and CASSACA grant CCJRF2105.
\end{acknowledgements}

%
%

\bibliography{ms2023-0340}
\bibliographystyle{raa}

\end{CJK*}
\end{document}